\documentclass[10pt, conference, letterpaper]{IEEEtran}
\IEEEoverridecommandlockouts

\usepackage{amsmath,amsfonts,graphicx,amssymb}
\usepackage{amsthm}
\usepackage{algorithm}
\usepackage{algorithmic}
\usepackage{booktabs}

\usepackage{subfigure} 

\usepackage{caption}
\usepackage{subcaption}
\usepackage{makecell}
\usepackage{xcolor}
\usepackage{comment}
\usepackage{soul}
\usepackage{multirow}
\usepackage{multicol}
\usepackage{cite}

\begin{document}

\title{Toward Scalable and Efficient Visual Data Transmission in 6G Networks}

\author{Junhao Cai, \textit{Student Member, IEEE}, Taegun An, \textit{Student Member, IEEE}, Changhee Joo, \textit{Senior Member, IEEE}
\thanks{C. Joo is the corresponding author.}
\thanks{This work was supported in part by the NRF grant funded by the Korea government (MSIT) (No. 2022R1A5A1027646).}
}

\maketitle 

\begin{abstract}
6G network technology will emerge in a landscape where visual data transmissions dominate global mobile traffic and are expected to grow continuously, driven by the increasing demand for AI-based computer vision applications. This will make already challenging task of visual data transmission even more difficult.
In this work, we review effective techniques for visual data transmission, such as content compression and adaptive video streaming, highlighting their advantages and limitations.
Further, considering the scalability and cost issues of cloud-based and on-device AI services, we explore distributed in-network computing architecture like fog-computing as a direction of 6G networks, and investigate the necessary technical properties for the timely delivery of visual data. 
\end{abstract}

\section{Introduction}

Advancements in smart devices and streaming services drastically increased mobile data transmission usage. As of 2020, video took two-thirds of global mobile data traffic and is expected to keep increasing~\cite{media_traffic}, which will be further accelerated with the introduction of AI-based computer vision applications. The rapid growth in data traffic creates significant challenges to existing network infrastructure and requires novel solutions to manage and optimize data transmission. Considering the traffic volume, efficient visual data transmission techniques will be the most crucial element in the 6G networks. 

To address the challenges of increasing data traffic, 6G may enhance conventional techniques and adopt innovative techniques. For example, terahertz communication can offer much higher bandwidth which is necessary for video streaming and other bandwidth-demanding applications~\cite{Terahertz}. Reconfigurable Intelligent Surfaces (RIS) can be used to adjust signal strength and manage interference~\cite{surfaces}. Edge computing can be used to significantly reduce latency in real-time video streaming~\cite{Vision}.
More directly related to visual data transmission, Dynamic Adaptive Streaming over HTTP (DASH)~\cite{dash_mmsys11} has been widely adopted for adaptive streaming and used to deliver video content over the Internet. In DASH, the server cuts video data into several chunks and pre-encodes each chunk with multiple bitrates. Based on the client's request and feedback, it streams video chunks of the best bitrate. With DASH, video streaming through the Internet has experienced less buffering and playback issues.

In this work, we overview several effective techniques for visual data transmissions with a focus on compression and adaptive transmission. Then we investigate the necessary properties of 6G networks for seamless delivery of visual data.


\section{Visual Data Compression for Transmission}

Dynamic network conditions and limited resources make it challenging to optimize overall network performance and accommodate the continuously increasing volume of visual data. Consequently, compressing visual data has become essential for optimizing network resources and enhancing the quality of delivered content. In this section, we introduce some conventional visual data compression frameworks and recent AI-based techniques, discussing their advantages and limitations.

Conventional visual data compression consists of three key modules: transformation, quantization, and entropy coding~\cite{benchmark}. During transformation, visual data is converted into compact and independent coefficients. The JPEG standard exploits the Discrete Cosine Transform (DCT) to convert 8×8 partitioned images, and JPEG 2000 employs the Discrete Wavelet Transform (DWT) on multi-resolution image representation to improve the coding performance and enable image conversion across scales. In the quantization process, less informative elements in the transformed coefficients are removed using methods like vector quantization or trellis-coded quantization. These independent and quantized coefficients are then compressed using entropy coding techniques such as Huffman coding, arithmetic coding, and context-adaptive binary arithmetic coding in various codecs, preparing them for network transmission.

There are well-known conventional codecs for image (BPG) and video (HEVC, VVC) compression that utilize intra-prediction and in-loop filtering to reduce spatial and inter-block redundancy, thereby improving reconstruction quality. However, further improving these codecs is challenging; end-to-end optimization of key modules often fails due to the complex dependencies between them, and enhancing the performance of a single module does not necessarily lead to overall performance improvement. Additionally, their block-based methods can cause blocking artifacts. 
These limitations can result in performance degradation, particularly when fine-tuning is required for adaptation to downstream AI tasks. Table~\ref{tab:classification_accuracy} shows the decrease in top-1 classification accuracy for several pre-trained models using the ILSVRC 2012 dataset compressed at a Bpp value of 0.5 with JPEG~\cite{Discernible}.


With the rapid development of deep learning, numerous studies have explored the potential of artificial neural networks (ANNs) for end-to-end optimized image compression frameworks. These learning-based methods are relatively free from the complex interdependencies between modules, facilitating the joint optimization of the entire framework and enabling partial improvements of individual modules to result in superior overall performance. 

A variational image compression framework using a scale hyperprior is proposed to capture spatial dependencies in latent representations and enhance rate-distortion performance~\cite{ballemshj18}. This framework employs a Gaussian distribution with trainable variance parameters as an entropy model, providing a more accurate estimation of the distribution of latent variables and thus achieving better compression efficiency. It has also achieved state-of-the-art performance on benchmark datasets compared to traditional codecs like JPEG and JPEG2000.
This framework is further extended by introducing autoregressive and hierarchical priors that better capture local and global dependencies, resulting in better entropy modeling and rate-distortion performance~\cite{minnenbt18}. Convolutional neural networks (CNNs) are used in the autoregressive and hierarchical components to learn complex dependencies in the data, outperforming both conventional and learned compression methods in image reconstruction tasks based on PSNR and MS-SSIM metrics.

Building upon these advancements and with the advent of the attention mechanism, a framework incorporating discretized Gaussian mixture likelihoods and attention modules has been proposed~\cite{cheng2020image}. This framework uses a mixture of Gaussians to model the distribution of latent variables more flexibly and accurately, while attention modules focus on more complex image patterns. Experimental results demonstrate that this approach achieves state-of-the-art performance, generating more visually plausible results with better compression quality compared to traditional codecs and other learned compression methods.

Although the aforementioned deep learning-based compression methods have structural advantages and achieve better performances, they face some limitations. The ANNs used in those frameworks require extensive training processes with large amounts of data and computational resources, leading to higher latency in actual implementation, which complicates their application in real-time services. In addition, due to the black-box nature of deep learning, learning-based compression methods typically struggle with understanding and diagnosing compression artifacts and errors compared to conventional methods.

\begin{table}[t]
    \centering
    \caption{Top-1 accuracy reductions on various models using JPEG compression, ILSVRC 2012 dataset, bpp = 0.5.}
    \label{tab:classification_accuracy}
    \begin{tabular}{cc}
        \toprule
        \textbf{Model} & \textbf{Reduction in Top-1 acc. (\%)} \\
        \midrule
        \textit{ResNet-18} & 7.3 \\
        \textit{ResNet-50} & 7.7 \\
        \textit{MobileNetV2} & 12.2 \\
        \textit{ShuffleNetV2} & 10.6 \\
        \textit{DenseNet169} & 7.6 \\
        \bottomrule
    \end{tabular}
\end{table}

\section{Adaptive Video Streaming}

For efficient video streaming, numerous research efforts have focused on  enhancing streaming quality through Adaptive Bitrate (ABR) algorithms, which adjust streaming bitrate according to available bandwidth. For instance, Festive~\cite{Festive} uses a harmonic mean to predict throughput, BBA and BOLA rely on buffer capacity to determine video bitrate, and Pensieve employs reinforcement learning to adapt the bitrate.
In video encoding, besides traditional methods like H.264/265, deep learning-based methods such as DVC and Elf-VC have gained significant attention. Scalable Video Coding (SVC)~\cite{SVC} can better adjust the coding rate according to network conditions due to its layered coding approach. Jigsaw focuses on minimizing encoding and decoding latency, thus offering lightweight layered coding strategies. Software-defined coding utilizes neural codecs, enabling more adaptable encoding tailored for specific video types or usage scenarios.

More recently, there have been a few notable works that adopted layered coding. Octopus~\cite{octopus_chen_arxiv22} allows applications to aggressively send data, with intermittent nodes in the network dropping packets in a prioritized manner according to their layers. This enables applications to specify how their content can be prioritized. When combined with a congestion control algorithm, it achieves high link utilization while minimizing queueing delays.
Another interesting scheme using layered coding is SWIFT~\cite{Swift}, which has developed a layered neural codec that avoids cross-layer compression overhead and includes a one-shot decoder capable of decoding any subset of these layered codes simultaneously. By eliminating the waiting time for high-priority layer content, it ensures timely delivery of high-quality content. 
Although these methods are innovative, they tend to overwhelm the network by injecting more packets, potentially leading to a waste of scarce bandwidth resources.

\section{Current Challenges and Proposed Solutions}

The aforementioned approaches require client feedback for bitrate control, introducing substantial feedback delays and inaccurate estimation of available resources at transmission time. Although Octopus relocates content adaptation from the server to network nodes, enhancing flexibility and responsiveness without client feedback, challenges such as video quality fluctuations, key frame loss, and resource allocation issues in multi-user environments persist.

Another significant drawback comes from the centralized processing architecture. Most schemes employ a client-server model, where the server performs extensive compression and the client decompresses the received codes~\cite{benchmark,ballemshj18,minnenbt18,cheng2020image}. This design places a substantial burden on the server, limiting its ability to adapt to client-side demands and often failing to scale as the number of clients increases. Modern communication networks incorporate substantial computing resources in a distributed manner, such as edge servers. Therefore, developing appropriate compression techniques that can accommodate the increasing and diverse client requests for video streaming will be of paramount importance in future network architectures.

To this end, we suggest necessary properties for scalable and efficient visual data transmission.

\begin{itemize} 
    \item \textbf{Distributed deployment:} It is important to develop visual compression models suitable for distributed environments, enabling compression at multiple stages. Each node in the proposed network would be capable of applying further compression based on its processing capacity and current network conditions. This approach leverages the distributed nature of modern networks, potentially reducing the burden on centralized servers.
    
    \item \textbf{Adaptive compression:} We suggest implementing algorithms that dynamically adjust compression levels based on real-time data characteristics. These mechanisms might regulate bitrate, resolution, and/or frame rate according to the network path from source to destination. This adaptive approach considers potential bandwidth variations and delays due to congestion, ensuring optimal performance across diverse network conditions.
    
    \item \textbf{Edge intelligence:} Lightweight AI solutions need to be integrated at the network edge to facilitate rapid decision-making and data processing. These edge-based models aim to contribute to distributed parallel computing, enhance response times, and ensure efficient utilization of network resources. By processing data closer to its source, this can potentially mitigate latency issues and improve overall system responsiveness.
\end{itemize}
Along with these properties, we can develop a robust framework for scalable and efficient visual data transmission, better suited to meet the demands of ever-increasing visual data traffic.

\section{Conclusion} 

In this paper, we explore the challenges and advancements in visual data transmission for upcoming 6G networks. We discuss the evolution from traditional compression methods to AI-based approaches, highlighting the potential of deep learning techniques to provide more flexible and efficient compression. Additionally, we examine various adaptive data transmission strategies, including recent innovations in layered coding approaches aimed at enhancing the quality of content delivery while optimizing network utilization.

Despite these promising developments, several challenges persist. They include feedback delays in adaptive streaming, quality fluctuations, and the limitations of centralized compression models. To address these issues, we consider several interesting properties: distributed deployment of compression models, adaptive compression algorithms, and edge intelligence. These properties leverage the distributed nature of modern networks, allowing dynamic adjustments based on network conditions and reducing reliance on centralized servers. 
As we progress towards 6G networks, future research should focus on refining these approaches and developing new technologies to meet the demanding requirements of next-generation visual data transmission. By achieving the discussed properties, we aim to create a robust framework for scalable and efficient visual data transmission, better suited to handle the ever-increasing volume of visual data traffic.

\section{Acknowledgment}

This work was supported by the NRF grant funded by the Korea government (MSIT) (No. 2022R1A5A1027646).



\end{document}